\documentclass[aps,twocolumn,showpacs,superscriptaddress]{revtex4-1}

\newcommand{\ifCJK}
  {\iffalse}

\ifCJK
\usepackage{CJK}
\fi

\usepackage{amsmath,bm,graphicx}
\usepackage{times}

\newcommand{\figsize}{8cm}
\newcommand{\fight}{8cm}

\newcommand{\arxiv}[1]{\textit{E-print} {arXiv:#1}}
\newcommand{\pofl}[3]{Phys.\ Fluids \textbf{#1}, {#2} (#3)}
\newcommand{\popl}[3]{Phys.\ Plasmas \textbf{#1}, {#2} (#3)}
\newcommand{\prll}[3]{Phys.\ Rev.\ Lett.\ \textbf{#1}, {#2} ({#3})}

\newcommand{\exb}{$\Vec{E} \times \Vec{B}$ }
\newcommand{\fig}[1]{Fig.\ \ref{#1}}
\newcommand{\figs}[2]{Figs.\ \ref{#1}--\ref{#2}}
\newcommand{\kpar}{k_{\parallel}}
\newcommand{\kperp}{k_{\perp}}

\newcommand{\gyroavg}[2]{\langle #1 \rangle_{\Vec{#2}}}

\newcommand{\sect}[1]{Sec.\ \ref{#1}}
\newcommand{\unit}[1]{\bm{e}_{#1}}
\renewcommand{\Vec}[1]{\bm{#1}}
\newcommand{\vperp}{v_{\perp}}
\newcommand{\vpar}{v_{\parallel}}
\newcommand{\vth}{v_\textrm{th}}

\newcommand{\FourHank}[1]{\hat{\breve{#1}}}
\newcommand{\Four}[1]{\hat{#1}}
\newcommand{\Hank}[1]{\breve{#1}}
\newcommand{\gk}{\Four{g}({\Vec{k}})}
\newcommand{\phik}{\Four{\varphi}(\Vec{k})}
\newcommand{\Norm}[1]{\tilde{#1}}

\newcommand{\UMD}{Department of Physics and IREAP, University of Maryland, College Park, Maryland 20742}

\begin{document}

\ifCJK
\begin{CJK}{UTF8}{}
\fi

\title{
  Freely decaying turbulence in two-dimensional electrostatic gyrokinetics
}

\ifCJK
\author{T.~Tatsuno (龍野 智哉)}
\else
\author{T.~Tatsuno}
\fi
\email{tomo.tatsuno@uec.ac.jp}
\affiliation{\UMD}
\affiliation{Department of Communication Engineering and Informatics,
  University of Electro-Communications, Chofu, Tokyo 182-8585, Japan}

\author{G.~G.~Plunk}
\affiliation{\UMD}
\affiliation{Max-Planck-Institut f\"{u}r Plasmaphysik, 17491 Greifswald,
  Germany}

\author{M.~Barnes}
\affiliation{Plasma Science and Fusion Center,
  Massachusetts Institute of Technology, Cambridge, Massachusetts 02139}

\author{W.~Dorland}
\affiliation{\UMD}

\author{G.~G.~Howes}
\affiliation{Department of Physics and Astronomy, University of Iowa,
  Iowa City, Iowa 52242}

\ifCJK
\author{R.~Numata (沼田 龍介)}
\else
\author{R.~Numata}
\fi
\affiliation{\UMD}
\affiliation{Graduate School of Simulation Studies, University of Hyogo,
  Kobe, Hyogo 650-0047, Japan}

\begin{abstract}
  In magnetized plasmas, a turbulent cascade occurs in phase space at scales smaller than the thermal Larmor radius (``sub-Larmor scales'') [Phys.\ Rev.\ Lett.\ \textbf{103}, 015003 (2009)].
  When the turbulence is restricted to two spatial dimensions perpendicular to the background magnetic field, two independent cascades may take place simultaneously because of the presence of two collisionless invariants.
  In the present work, freely decaying turbulence of two-dimensional electrostatic gyrokinetics is investigated by means of phenomenological theory and direct numerical simulations.
  A dual cascade (forward and inverse cascades) is observed in velocity space as well as in position space, which we diagnose by means of nonlinear transfer functions for the collisionless invariants.
  We find that the turbulence tends to a time-asymptotic state, dominated by a single scale that grows in time.
  A theory of this asymptotic state is derived in the form of decay laws.
  Each case that we study falls into one of three regimes (weakly collisional, marginal, and strongly collisional), determined by a dimensionless number $D_*$, a quantity analogous to the Reynolds number.
  The marginal state is marked by a critical number $D_* = D_0$ that is preserved in time.
  Turbulence initialized above this value become increasingly inertial in time, evolving toward larger and larger $D_*$; turbulence initialized below $D_0$ become more and more collisional, decaying to progressively smaller $D_*$.
\end{abstract}


\maketitle
\ifCJK
\end{CJK}
\fi

\section{Introduction}

Plasma turbulence plays important roles in fusion devices and various space and astrophysical situations, where it is an essential phenomenon underlying transport of mean quantities and particle heating \cite{WatanabeSugama-PoP04, Hennequin-PPCF04, Idomura-PoP06, Candy-PPCF07, Garbet-NF10, Bale-PRL05, Howes-PRL08, GarySaitoLi, Sahraoui-PRL09, Chen-PRL10, Schekochihin-ApJS09}.
For these collisionless or weakly collisional plasmas, such turbulence requires a kinetic description in phase space, especially at small scales where dissipation takes place.

Turbulence theory in kinetic phase space is more than a simple extension of Navier-Stokes turbulence into higher dimensions as velocity space is not exactly equivalent to position space:
For instance, there is no translational symmetry (i.e.\ small velocities are not equivalent to large velocities), and there is always some large-scale velocity dependence imposed by the background distribution function (e.g.\ on the scale of the thermal velocity for a Maxwellian background); see also Ref.\ \cite{Plunk-Big-Fjortoft}, Sec.\ 2.1.
However, in some simplified cases, classical fluid dynamical theories \cite{Kolmogorov, Fjortoft-Tellus53} can be naturally extended into phase space \cite{Schekochihin-PPCF08, Schekochihin-ApJS09, Plunk-JFM10, Tatsuno-PRL09, Tatsuno-JPFRS10, Plunk-Fjortoft}.
In magnetized plasmas, the gyrokinetic (GK) theory \cite{Catto-PP78, AntonsenLane-PoF80, FriemanChen-PoF82, Howes-ApJ06} provides the minimal kinetic description of the low-frequency turbulence.

In electrostatic gyrokinetics, nonlinear interactions introduce a cascade of perturbed entropy (which is proportional to the perturbed free energy at the sub-Larmor scales) to smaller scales both in position and velocity space \cite{Schekochihin-PPCF08, Schekochihin-ApJS09, Plunk-JFM10, Tatsuno-PRL09, Tatsuno-JPFRS10, Plunk-Fjortoft}.
When the turbulence is restricted to two position-space dimensions, the system has two collisionless invariants.
One of them is the free energy or entropy which is also an invariant in three dimensions (3D), and another, approximately related to kinetic energy in the long wave-length limit, is particular to two dimensions (2D).
In this regard, 2D gyrokinetic turbulence is analogous to 2D fluid turbulence, and indeed, reduces to it in a particular long wave-length limit \cite{TaylorMcNamara-PoF71, Plunk-JFM10}.
These two invariants cannot share the same local-interaction space in a Kolmogorov-like phenomenology, which leads to a dual cascade (forward and inverse cascades) \cite{Idomura-PoP06, Fjortoft-Tellus53, Kraichnan-PoF67, HasegawaMima-PoF78, Horton-PoP00}.
As the nonlinear term of 2D gyrokinetics is identical in form to that of 3D gyrokinetics, the understanding of the nonlinear interaction in purely 2D system will serve as a foundation for understanding general 3D magnetized plasmas \cite{Plunk-Fjortoft}.

In this paper we focus on the freely decaying turbulence problem for the electrostatic 2D GK system.
We first introduce the GK equation briefly and describe its basic nature in \sect{sec:equations}.
We also review basic characteristics of the nonlinear phase mixing at sub-Larmor scales that are reported in Refs.\ \cite{DorlandHammett-PoFB93, Tatsuno-PRL09, Tatsuno-JPFRS10}.
Sections \ref{sec: theory} and \ref{sec:simulation} are the main contents of the paper.
In Sec.\ \ref{sec: theory}, we describe a phenomenological theory of the dual cascade in freely decaying turbulence based on the theory first developed for 2D fluid turbulence \cite{Chasnov-PoF97}.
There are three regimes for the turbulence:
These correspond to weakly collisional, marginal, and strongly collisional cases.
In the strongly collisional case, dissipation acts strongly and the system decays to become more and more dissipative.
In the marginal case, the collisional dissipation balances with nonlinear (inertial) turnover and the system is preserved in this state.
For weakly collisional cases, the turbulence is, in a sense, able to escape the effects of dissipation, becoming less and less collisional in time, and tending asymptotically to a state of zero electrostatic energy decay.
Section \ref{sec:simulation} presents the results of numerical simulation of the freely decaying turbulence.
We demonstrate the inverse (forward) transfer of energy (entropy) in the direct numerical simulation and investigate the time-asymptotic decay laws of the two collisionless invariants, comparing with the theory developed in Sec.\ \ref{sec: theory}.
We conclude with a summary of our results in \sect{sec:summary}.

\section{Phase-space turbulence}
\label{sec:equations}

\subsection{Gyrokinetic equations and invariants}

We first introduce the gyrokinetic (GK) model briefly \cite{Catto-PP78, AntonsenLane-PoF80, FriemanChen-PoF82, Howes-ApJ06}.
Since we are concerned with turbulence in magnetized plasmas, the dynamics of interest is much slower than particle gyromotion.
The gyromotion is thus averaged over, eliminating gyroangle dependence from the system.
The GK system has 3 spatial coordinates ($x$, $y$, $z$), and 2 velocity coordinates ($\vperp$, $\vpar$), where $\perp$ and $\parallel$ denote perpendicular and parallel directions to the background magnetic field, respectively.
We assume the background plasma and magnetic field are uniform in space and time.
It is necessary to distinguish between the particle coordinate $\Vec{r}$ and the gyrocenter coordinate $\Vec{R}$.
These coordinates are connected by the Catto transform \cite{Catto-PP78}
\begin{equation}
  \Vec{R} = \Vec{r} + \frac{\Vec{v} \times \unit{z}}{\Omega},
  \label{Catto transform}
\end{equation}
where $\unit{z}$ is the unit vector along the background magnetic field and $\Omega$ is the gyrofrequency.

We further reduce the GK equation to 2D in position space, or 4D in phase space \cite{4D simulation}, by ignoring variation along the mean field ($\kpar=0$).
This not only reduces the dimension of the system but also removes one of the mechanisms of creating velocity-space structure --- linear parallel phase mixing (Landau damping), a much more familiar and better understood phenomenon than the nonlinear perpendicular phase mixing, on which we will concentrate in this paper.
The resulting GK equation (for the ions) is
\begin{equation}
  \label{eq:gk}
  \frac{\partial g}{\partial t} +  \frac{c}{B_0} \{ \gyroavg{\varphi}{R},
    g \} = \gyroavg{C}{R},
\end{equation}
where $\gyroavg{\cdot}{R}$ is the gyroaverage holding the guiding center position $\Vec{R}$ constant, $g = \gyroavg{\delta f}{R}$ is the gyroaverage of the perturbed ion distribution function $\delta f$, $\varphi$ is the electrostatic potential, $B_0$ is the background magnetic field (aligned with the $z$-axis), and $\{ f, g \}= \unit{z} \cdot (\nabla f \times \nabla g)$.
The collision operator $C$ we use in our simulations describes pitch-angle scattering and energy diffusion with proper conservation properties \cite{Abel-PoP08, Barnes-PoP09} (see also Appendix \ref{sec: collision}).

The potential $\varphi$ in Eq.\ \eqref{eq:gk} is calculated from the quasineutrality condition:
Written in the Fourier space, it is
\begin{equation}
  \label{eq:qn}
  \frac{n_0 q_i^2}{T_{0i}} (1 + \tau - \varGamma_0) \phik = q_i \int
    J_0 \left( \frac{\kperp \vperp}{\Omega}\right) \gk
    \, d\Vec{v},
\end{equation}
where the hat denotes the Fourier coefficients, $J_0$ is the Bessel function, representing, in Fourier space, the gyroaverage at fixed particle position, $\varGamma_0 = I_0(b) e^{-b}$, $I_0$ is the modified Bessel function of the first kind, $b = \kperp^2 \rho^2 / 2$, $\rho$ is the ion thermal Larmor radius, $q$ is the charge, $n_0$ and $T_0$ are the density and temperature of the background Maxwellian $F_0$, and $i$ and $e$ are the species indices.
One may use $\tau = - q_e T_{0i} / q_i T_{0e}$ for Boltzmann-response (3D) electrons or $\tau = 0$ for no-response (2D) electrons \cite{Plunk-JFM10}.
Hereafter we use the no-response electrons as in Refs.\ \cite{Tatsuno-PRL09, Tatsuno-JPFRS10} because formally, the electrons cannot contribute to the potential if $\kpar = 0$ exactly \cite{DorlandHammett-PoFB93}.
This choice is not very important as it only introduces minor differences in various prefactors.

The 2D electrostatic GK system possesses two quadratic positive-definite collisionless invariants \cite{Plunk-JFM10},
\begin{align}
  \label{eq:wg}
  W &= \sum_{\Vec{k}} \int \frac{T_{0i} |\Four{g}(\Vec{k})|^2}{2 F_0} \, d\Vec{v}, \\
  \label{eq:esi}
  E &= \frac{n_0 q_i^2}{2T_{0i}} \sum_{\Vec{k}} (1 - \varGamma_0) |\phik|^2.
\end{align}
There are various ways to choose two independent invariants in our system.
In Refs.\ \cite{Tatsuno-PRL09, Tatsuno-JPFRS10} total perturbed entropy (or free energy), $W_{\rm tot} = W + E$ (see Eq.\ (3.9) of Ref.\ \cite{Plunk-JFM10}), is used in order to make the connection to thermodynamics.
Here we use the quantity $W$ for the sake of simplicity \cite{Plunk-JFM10}.
In fact, $g^2$ averaged over $\Vec{R}$ is itself conserved (that is, it is conserved for each value of $\vperp$) \cite{Plunk-JFM10, ZhuHammett-PoP10}.
However, it is sufficient for the purposes of this paper to consider only the integrated quantity $W$.
One can adapt the arguments of Ref.\ \cite{Fjortoft-Tellus53} to show that the presence of conserved quantities $W$ and $E$ implies a dual cascade (i.e., both forward and inverse cascades) \cite{Plunk-Fjortoft}.

\subsection{Nonlinear phase mixing}

\begin{figure}[bt] 
  \centerline{
    \includegraphics[width=5cm]{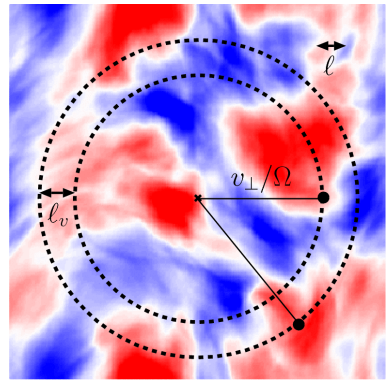}
  }
  \caption{
    (Color online)
    Schematic view of the nonlinear phase mixing:
    when the fluctuation scale $\ell$ is comparable to
    or smaller than the Larmor radius $\rho$, the gyroaverage of the 
    electric field induces a decorrelation of the distribution function at the 
    velocity-space scale corresponding to the difference in Larmor radii
    $\ell_v=\delta \vperp/\Omega\sim\ell$.
    This figure is reproduced with permission from Ref.\ \cite{Tatsuno-PRL09}.
    Copyright (2009) by the American Physical Society
    (http://link.aps.org/abstract/PRL/v103/e015003).
  }
  \label{fig:mixing}
\end{figure}

Due to the neglect of the parallel streaming term, the creation of velocity space structure originates solely from the advection of the distribution function by the gyroaveraged \exb drift [the nonlinear term in Eq.\ \eqref{eq:gk}].
For small-scale electric fields, particles with different gyroradii execute different \exb motions because they ``see'' different effective potentials; this leads to nonlinear phase mixing and other novel phenomena \cite{Schekochihin-PPCF08, Schekochihin-ApJS09, DorlandHammett-PoFB93, Gustafson-PoP08}.
As the turbulence cascades through phase-space, the excitation of fluctuations at spatial scale $\ell$ induces velocity structure of scale $\delta \vperp$ in the perpendicular velocity space which corresponds to the difference of the Larmor radii $\ell_v = \delta \vperp/\Omega \sim \ell$ (see \fig{fig:mixing} and Refs.\ \cite{Schekochihin-PPCF08, Schekochihin-ApJS09, Plunk-JFM10, Tatsuno-PRL09, Tatsuno-JPFRS10}).
In other words, when the spatial decorrelation scale is $\ell$, two particles with Larmor radii separated by $\ell_v$ become decorrelated, since the gyroaveraged potentials these two particles feel are different.
This nonlinear phase-space mixing effect was first pointed out by Dorland and Hammett \cite{DorlandHammett-PoFB93}.

In Refs.\ \cite{Tatsuno-PRL09, Tatsuno-JPFRS10} numerical simulations focused on the forward cascade of entropy, and showed a simultaneous creation of structures in position and velocity space in accordance with the theoretical prediction \cite{Schekochihin-PPCF08, Schekochihin-ApJS09, Plunk-JFM10}.
\begin{figure} 
  \centerline{
    \includegraphics[height=5.5cm]{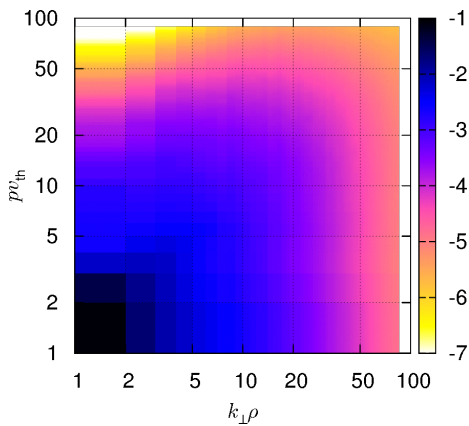}
  }
  \caption{
    (Color online)
    Two-dimensional spectral density
    $\log_{10} [\FourHank{W}(\kperp,p)/W_{\rm tot}]$ from one of the forward
    cascade simulation reported in Ref.\ \cite{Tatsuno-JPFRS10}.
    Kinetic turbulence proceeds in the position and velocity
    space simultaneously.
    This figure is reproduced with premission from Ref.\ \cite{Tatsuno-JPFRS10}.
    Copyright (2010) by the Japan Society of Plasma Science and Nuclear
    Fusion Research.
  }
  \label{fig: forward cascade result}
\end{figure}
Figure \ref{fig: forward cascade result} shows a normalized, time-averaged spectral density of entropy
\begin{equation}
  \FourHank{W}(\kperp,p) = \sum_{|\Vec{k}| = \kperp} p |\FourHank{g}(\Vec{k},p)|^2,
  \label{eq: 2d-spectrum}
\end{equation}
in a forward cascade simulation, where the $\vperp$ structure is characterized by the Hankel transform \cite{Plunk-JFM10, Hankel transform}
\begin{equation}
  \FourHank{g}(\Vec{k},p) = \int J_0(p\vperp) \Four{g}(\Vec{k}, \vperp, \vpar)
    \, d\Vec{v},
  \label{eq: Hankel transform}
\end{equation}
and $\FourHank{}$ denotes the Fourier-Hankel coefficients.

In analogy with the Reynolds number in fluid turbulence, we may introduce an amplitude-dependent dimensionless number $D$ \cite{Tatsuno-PRL09}, the ratio of collision time to nonlinear decorrelation time measured at the thermal Larmor radius, which characterizes the smallest scales created in both position and velocity space by $D^{-3/5}$.
Thus, $D$ quantifies both how ``inertial'' the turbulence is (in the sense of the Reynolds number) and how ``kinetic'' it is, because it measures the nonlinear turnover at the thermal Larmor radius, which marks the beginning of the ``nonlinear phase-mixing range.''
The degree of freedom, corresponding to computational problem size, may also be characterized by $D$, and scales as $D^{9/5}$ in three phase-space dimensions, consisting of two position-space and one velocity-space dimensions.

We note here that the statistical description of our phase-space turbulence requires a 2D spectral space $(\kperp, p)$, in contrast to isotropic fluid turbulence, which requires only the scalar wave number.
We will refer to this 2D spectrum $\FourHank{W}$ in the following sections.

\section{Theory of freely decaying turbulence}
\label{sec: theory}

\subsection{Dual cascade}
\label{sec:phenomenology}

\begin{figure} 
  \centerline{
    \includegraphics[width=6cm]{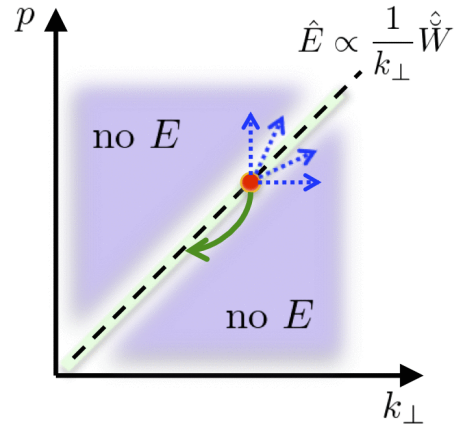}
  }
  \caption{(Color online)
    Schematic view of the freely decaying turbulence in the $(\kperp,p)$ space.
    A diagonal Fourier-Hankel mode (indicated by the red circle) cascades towards small scales in position and velocity space according to the forward cascade of $W$ (blue dotted arrows).
    Nonlinear interaction conserves both $W$ and $E$, so the forward cascade must be accompanied by the excitation of larger-scale modes on the diagonal (green solid arrow), which corresponds to the inverse cascade of $E$.
  }
  \label{fig:phenom}
\end{figure}

With the use of the Hankel transform \eqref{eq: Hankel transform} for velocity space and the conventional Fourier decomposition for position space, we may discuss the evolution of Fourier-Hankel modes of $g$ in $(\kperp,p)$ space.
Figure \ref{fig:phenom} depicts the simplest example of the evolution of $\FourHank{W}(k_{\perp},p)$ in the freely decaying turbulence.

By applying the definition \eqref{eq: Hankel transform} to Eq.\ \eqref{eq:qn}, we find that the Fourier-Hankel modes $\FourHank{g}(\Vec{k},p)$ with $\kperp\rho \neq p\vth$ (purple shaded regions marked by ``no $E$'' in \fig{fig:phenom}) give no contribution to $\phik$ due to the orthogonality of the Bessel functions.
In these regions, therefore, fluctuations associated with this part of $W$ can in principle cascade to small scales with no effect on the spectral distribution of $E$.

On the other hand, the spectrum of $E$ satisfies \cite{Plunk-JFM10}
\begin{align}
  \Four{E}(\kperp) &:= \frac{n_0 q_i^2}{2T_{0i}}
    \sum_{|\Vec{k}| = \kperp} |\phik|^2 \nonumber \\
  &= \frac{T_{0i} \vth}{2 n_0 (1-\varGamma_0)^2 \kperp \rho} \FourHank{W}
    \left(\kperp, \frac{\kperp \rho}{\vth} \right), 
  \label{def: E spectrum}
\end{align}
where $\FourHank{W}$ is given by Eq.\ \eqref{eq: 2d-spectrum}, i.e., while $W$ is the sum over the entire $(\kperp,p)$-plane, the second invariant $E$ is only composed of the ``diagonal'' $\kperp \rho = p \vth$ components (dashed line in \fig{fig:phenom}).
In the small-scale limit ($\kperp\rho \gg 1$), we may approximate $\varGamma_0 \sim 1/(\kperp\rho)$, so Eq.\ \eqref{def: E spectrum} implies
\begin{equation}
  \Four{E}(\kperp) \propto \frac{1}{\kperp}
    \FourHank{W}\left(\kperp,\frac{\kperp \rho}{\vth}\right),
  \label{E W relationship}
\end{equation}
which is indicated for the diagonal components in \fig{fig:phenom}.

As we will see, the decaying turbulence evolves to a state of local cascade.
As a consequence of Eq.\ \eqref{E W relationship}, this state is characterized by an inverse cascade of $E$ and forward cascade of $W$, which can be explained as follows \cite{Plunk-Fjortoft}.

The energy of a diagonal Fourier-Hankel mode (indicated by the red circle) cascades towards small scales in position and velocity space according to the forward cascade of $W$ (blue dotted arrows).
Nonlinear interaction conserves both $W$ and $E$, so the forward cascade must be accompanied by the excitation of larger-scale modes on the diagonal (green solid arrow), which corresponds to the inverse cascade of $E$.
Simulation results reported in Sec.\ \ref{sec:simulation} indicate that this is the universal property of the time-asymptotic state of freely decaying turbulence.
However, there are other interactions possible in the transient and/or driven cases (for more details on such processes, see Refs.\ \cite{Plunk-Fjortoft,Plunk-Big-Fjortoft}).

\subsection{Decay laws}
\label{sec:decay law}

In this section, we derive scaling laws for the decay of collisionless invariants based on some simple phenomenological arguments.

The underlying assumptions are as follows.
First, we assume that collisionless invariants are dominated by a single scale $l_*$, in the same way energy in fluid turbulence is dominated by the ``energy-containing scale.''
Then, in terms of the amplitude at this scale,
\begin{align}
  &W \sim \frac{T_{0i}\vth^6}{n_0} g_*^2,
  &E \sim \frac{n_0 q_i^2}{T_{0i}} \varphi_*^2,
  \label{invariants dominated by a single scale}
\end{align}
where $g_*$ and $\varphi_*$ are the rms values of the distribution function and potential associated with the scale $l_*$.
Second, we assume that the evolution of $l_*$ is governed by the inverse cascade along the diagonal $\kperp \rho \sim p \vth$, so defining $k_* := 1/l_*$, we have
\begin{equation}
  W \sim k_* \rho E.
  \label{invariants ratio}
\end{equation}
Both assumptions are found to be valid in the numerical simulations described in Sec.\ \ref{sec:simulation}.

Depending on the strength of collisions compared to that of turbulent dynamics, we may derive several different scaling laws.
In order to quantify the collisionality, we characterize the instantaneous turbulent state via a sub-Larmor version of the dimensionless number introduced in Ref.\ \cite{Tatsuno-PRL09}:
It is the ratio of the collision time to the nonlinear decorrelation time $\tau_*$ at the scale $l_*$ \cite{dstar},
\begin{equation}
  D_* = \frac{1}{\nu k_*^2 \rho^2 \tau_*}
    \sim \frac{\Omega E}{\nu (n_0T_{0i}W)^{1/2}}.
  \label{microscopic Dorland number}
\end{equation}
We have taken the collisional decay rate to scale as $\nu k_*^2 \rho^2$ because the GK collision operator is second order in velocity and spatial derivatives \cite{Abel-PoP08, Barnes-PoP09} and $\kperp \rho \sim p \vth$.
Note that in going from the second to the third expression in Eq.\ \eqref{microscopic Dorland number}, we used Eq.\ \eqref{invariants ratio} and
\begin{equation}
  \tau_*^{-1} \sim \frac{c}{\rho^{1/2}B_0} k_*^{3/2} \varphi_*,
  \label{nonlinear time}
\end{equation}
valid in the $k_* \rho \gg 1$ regime from the form of the convective derivative $(c/B_0) \nabla \gyroavg{\varphi}{R} \cdot \nabla$.
Note that the factor of $(k_*\rho)^{-1/2}$ is introduced due to the large argument expansion of the Bessel function associated with the gyroaverage of $\varphi_*$.
In analogy with the microscopic Reynolds number \cite{Chasnov-PoF97}, the initial value and the time evolution of $D_*$ provide a natural way to classify various physical regimes.

In the following paragraphs, we describe three different decay laws, classified using Eq.\ \eqref{microscopic Dorland number}.
They are the weakly collisional ($D_* \gg D_0$), marginal ($D_* \sim D_0$) and strongly collisional ($D_* \ll D_0$) cases.
Here the constant $D_0$ denotes the value of $D_*$ that divides these three regimes, which corresponds to the kinetic version of the ``critical Reynolds number'' in Ref.\ \cite{Chasnov-PoF97}.
$D_0$ is a universal constant by conjecture, which applies to the asymptotic state of the freely decaying turbulence as we explain below (see also \sect{sec: num decay law} and \fig{fig: dorland number}).

\paragraph{Weakly  collisional case}
In the asymptotic limit where collision frequency becomes negligible ($\nu \to 0$), the second invariant $E$ does not decay at all as it is transferred to larger scale where dissipation is inactive.
On the other hand, the first invariant $W$ is transferred to smaller scales, and is dissipated by small but finite collisions there.
The decay rate of $W$ is determined by the rate of its transfer to small scales, thus
\begin{align}
  &\frac{dE}{dt} \sim 0,
  &\frac{dW}{dt} \sim -\frac{W}{\tau_*},
\end{align}
where we used $\tau_*$ for the characteristic time of nonlinear transfer at scale $l_*$ [see Eq.\ \eqref{nonlinear time}].
Then from $\tau_* \sim t$ and $E \propto \varphi_*^2 \propto t^0$, we obtain
\begin{align}
  &E \sim {\rm const}.,
  &W \propto k_* \propto t^{-2/3},
  \label{asymptotic decay law}
\end{align}
where we used Eqs.\ \eqref{invariants ratio} and \eqref{nonlinear time}.
In this limit Eq.\ \eqref{microscopic Dorland number} implies that $D_*$ increases in time as $D_* \propto t^{1/3}$.

\paragraph{Marginal case}
As collision frequency becomes large, collisional damping at scale $l_*$ becomes important.
When it balances with the nonlinear transfer, the two terms in
\begin{equation}
  \frac{dW}{dt} \sim - \frac{W}{\tau_*} - \nu k_*^2 \rho^2 W,
  \label{entropy evolution for marginal case}
\end{equation}
become comparable, where we have taken the collisional decay rate to scale as $\nu k_*^2 \rho^2$.
From Eqs.\ \eqref{invariants dominated by a single scale}, \eqref{nonlinear time} and $\tau_* \sim t$, we obtain the following decay laws of collisionless invariants
\begin{align}
  &E \propto k_* \propto t^{-1/2},
  &W \propto t^{-1}.
  \label{marginal decay law}
\end{align}
Substitution of Eq.\ \eqref{marginal decay law} into Eq.\ \eqref{microscopic Dorland number} immediately leads to a constant $D_*$ which equals the marginal dimensionless number $D_0$.
It is noted that Eq.\ \eqref{entropy evolution for marginal case} implies that the second invariant also decays by collisions in a consistent manner:
\begin{equation}
  \frac{dE}{dt} \sim - \nu k_*^2 \rho^2 E.
  \label{energy evolution for dissipative case}
\end{equation}

\paragraph{Strongly collisional case}
In this case we may regard the turbulence to be fairly dissipative.
We find that $W$ and $E$ do not individually satisfy decay laws as powers of $t$.
However, Eq.\ \eqref{energy evolution for dissipative case} applies, as does the analogous equation describing the collisional decay of $W$,
\begin{equation}
  \frac{dW}{dt} \sim - \nu k_*^2 \rho^2 W.
\end{equation}
These equations imply [with the help of Eq.\ \eqref{invariants ratio}] decay laws for both $k_*$ and the ratio $E/W$:
\begin{align}
  &k_* \propto t^{-1/2},
  &\frac{E}{W} \propto t^{1/2}.
  \label{collisional decay law}
\end{align}
In this case we can deduce that $D_*$ decays in time.
It is noted that although Eq.\ \eqref{collisional decay law} holds for both marginal and strongly collisional cases, the individual decay laws for $W$ and $E$ may vary with $D_*$ [i.e.\ Eq.\ \eqref{marginal decay law} is not satisfied here]; however, the decay law for the ratio $E/W$ is robust.

\section{Simulation results}
\label{sec:simulation}

In this section we show the results of numerical simulations performed using the MPI-parallelized nonlinear gyrokinetic code AstroGK \cite{Numata-JCP10}.
All simulations are made in two spatial dimensions ($x$ and $y$) and two velocity dimensions (energy $\varepsilon = v^2$ and pitch angle $\lambda = \vperp^2 / \varepsilon$) \cite{4D simulation}.
The system size is restricted to $L_x = L_y = 2 \pi \rho$, so as to focus on the sub-Larmor regime.
Time is normalized by the initial turnover time
\begin{equation}
  \tau_{\rm init} = \frac{2 \pi B_0}{c k_0^2 ||\gyroavg{\varphi_{\rm init}}{R}||},
\end{equation}
where $||\gyroavg{\varphi}{R}|| = [(1/n_0) \iint |\gyroavg{\varphi}{R}|^2 F_0 \, d\Vec{R} \, d\Vec{v}]^{1/2}$, and $k_0$ is the wave number at which initial spectrum is peaked [see Eqs.\ \eqref{coherent init cond} and \eqref{random init cond} below].
Our biggest run (Run D in Table \ref{run table}) used 9,216 processor cores for about 50 wall-clock hours.

\subsection{Initial conditions}

In order to investigate the freely decaying turbulence, we prepared initial conditions peaked at a high wave number $k_0$ ($k_0 \rho \gg 1$) and made a series of simulations for varying initial wave number $k_0$ and collision frequency $\nu$.
Six runs were made for two kinds of velocity distribution functions (described below) and are listed in Table \ref{run table}.
\begin{table}
  \caption{
    Index of the runs described in \sect{sec:simulation}.
  }
  \begin{center} \begin{tabular}{ccccccc}
      \hline \hline
      Run & $\nu \tau_{\rm init}$ && $k_0 \rho$ & $N_x \times N_y$ &
        $N_{\varepsilon} \times 2N_{\lambda}$ & init.\ cond.\ 
        \\ \hline
      A & $1.7 \times 10^{-3}$ && $15$ & $64^2$ & $48^2$ & Eq.\ \eqref{coherent init cond} \\
      B & $4.2 \times 10^{-4}$ && $25$ & $128^2$ & $96^2$ & Eq.\ \eqref{coherent init cond} \\
      C & $3.3 \times 10^{-4}$ && $25$ & $128^2$ & $96^2$ & Eq.\ \eqref{coherent init cond} \\
      D & $5.2 \times 10^{-5}$ && $40$ & $256^2$ & $192^2$ & Eq.\ \eqref{coherent init cond} \\
      E & $4.2 \times 10^{-4}$ && $40$ & $256^2$ & $192^2$ & Eq.\ \eqref{random init cond} \\
      F & $3.8 \times 10^{-4}$ && $25$ & $256^2$ & $192^2$ & Eq.\ \eqref{random init cond} \\
      \hline \hline
    \end{tabular} \end{center}
  \label{run table}
\end{table}
The different initial velocity distributions allow us to vary the initial ratio of invariants.
For each initial velocity distribution, we made weakly and strongly collisional simulations.

\setcounter{paragraph}{0}
\paragraph{Coherent velocity distribution (Runs A--D)}
The first type of the initial velocity distribution is Bessel-like, with a Maxwellian envelope $F_0$,
\begin{equation}
  \Four{g}(\Vec{k},\vperp,\vpar) = g_0 \frac{\kperp^2}{k_0^2} \exp \left[
    - \left( \frac{\kperp-k_0}{k_{\rm w}} \right)^2 \right]
  J_0\left( \frac{\kperp \vperp}{\Omega} \right) F_0,
  \label{coherent init cond}
\end{equation}
where the width of the wave-number peak is $k_{\rm w} \rho = 1$.
From quasi-neutrality [Eq.\ \eqref{eq:qn}], it can be deduced that small-scale velocity oscillation whose period in velocity space is comparable to $\vth/(k_0\rho) = \Omega/k_0$ are needed to produce a finite potential (see the discussions in \sect{sec:phenomenology}).
Indeed, such oscillatory structure can be found in the eigenmodes of the entropy mode \cite{Ricci-PoP06}, which can be unstable at quite large $\kperp \rho$.
The Fourier-Hankel spectral density corresponding to Eq.\ \eqref{coherent init cond} is concentrated at $\kperp \rho = p \vth = k_0\rho$ as plotted in the left panel of \fig{fig: coherent 2d-spectra} corresponding to $t/\tau_{\rm init}=0$, and represented by the red spot in \fig{fig:phenom}.
For Eq.\ \eqref{coherent init cond} the initial ratio of the invariants is
\begin{equation}
  \frac{W_{\rm init}}{E_{\rm init}} \sim k_0 \rho,
  \label{coherent init ratio}
\end{equation}
which we note is the same as what is predicted for the time-asymptotic state, given by Eq.\ \eqref{invariants ratio}.

\paragraph{Random velocity distribution (Runs E and F)}
The second type is a random velocity distribution that represents velocity scales $\delta v / \vth \gtrsim 1 / (k_0 \rho)$,
\begin{align}
  \Four{g}(\Vec{k},\vperp,\vpar) &= g_0 \frac{\kperp^2}{k_0^2} \exp \left[
    - \left( \frac{\kperp-k_0}{k_{\rm w}} \right)^2 \right] \nonumber \\
  &\times \frac{1}{N} \sum_{j=1}^N
  (2 \delta_j - 1) \sqrt{p_j\vth} J_0 ( p_j \vperp ) F_0,
  \label{random init cond}
\end{align}
where the width of the wave-number peak is $k_{\rm w} \rho = 1$, $p_j\vth = (\kperp + k_{\rm w}) \rho \eta_j$, $\eta_j$ and $\delta_j$ are homogeneous random numbers in $(0,1)$, and $N=50$ is the number of random modes for each $\kperp$.
The factor of $\sqrt{p_j\vth}$ is introduced to cancel the same factor of the Bessel function in the asymptotic regime ($p_j \vth \gg 1$).
As is shown in the left panel ($t/\tau_{\rm init}=0$) of \fig{fig: random 2d-spectra}, Eq.\ \eqref{random init cond} corresponds to a high-density band parallel to the $p$ axis in the $(\kperp,p)$ spectral space.
In this case the initial ratio of the invariants is
\begin{equation}
  \frac{W_{\rm init}}{E_{\rm init}} \sim k_0^2 \rho^2.
  \label{random init ratio}
\end{equation}
Note that while Eq.\ \eqref{coherent init cond} is used to represent the velocity structure of a coherent mode, the distribution \eqref{random init cond} is designed to mimic the random nature of the velocity space that develops from forward cascade (see Ref.\ \cite{Tatsuno-PRL09}).

\subsection{Spectral evolution}

\begin{figure*} 
  \centerline{
    \includegraphics[width=16cm]{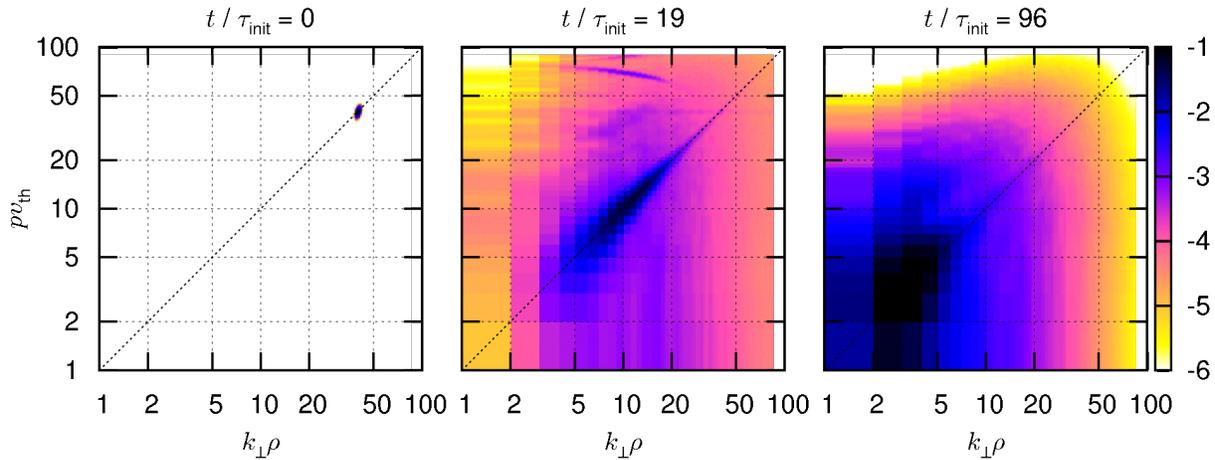}
  }
  \caption{(Color online)
    Time evolution of the 2D spectra $\log_{10} [\FourHank{W}(\kperp,p) / W]$
    for Run D [see Eq.\ \eqref{eq: 2d-spectrum} and Table \ref{run table}].
    Diagonal components ($\kperp\rho = p\vth$) are indicated by dotted lines.
  }
  \label{fig: coherent 2d-spectra}
\end{figure*}
\begin{figure*} 
  \centerline{
    \includegraphics[width=16cm]{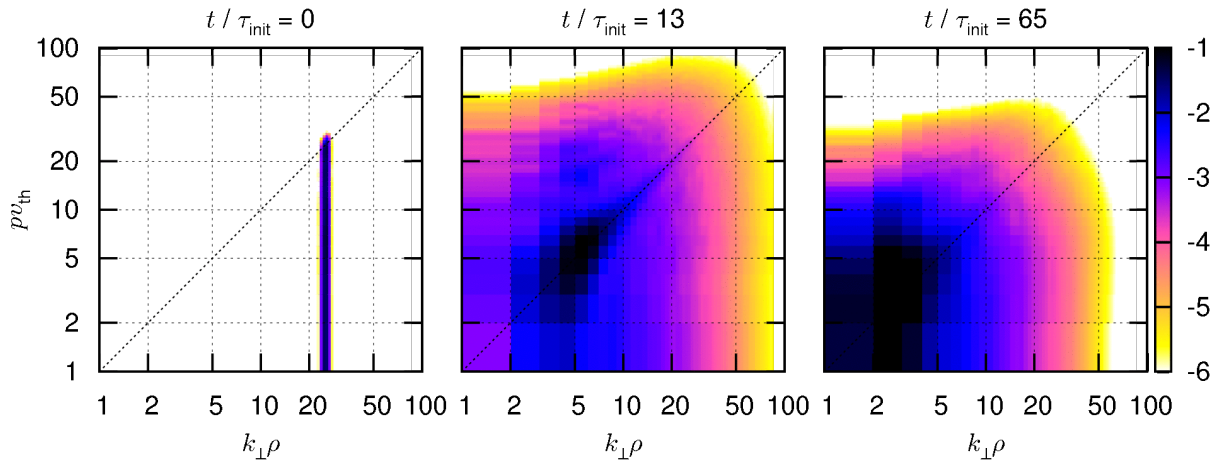}
  }
  \caption{(Color online)
    Time evolution of the 2D spectra $\log_{10} [\FourHank{W}(\kperp,p) / W]$
    for Run F [see Eq.\ \eqref{eq: 2d-spectrum} and Table \ref{run table}].
    The diagonal ($\kperp\rho = p\vth$) is indicated by dotted lines.
  }
  \label{fig: random 2d-spectra}
\end{figure*}
The time evolution of the 2D spectrum $\FourHank{W}(k_{\perp},p)$ [see Eq.\ \eqref{eq: 2d-spectrum}] for Runs D and F (Table \ref{run table}) is shown in Figs.\ \ref{fig: coherent 2d-spectra} and \ref{fig: random 2d-spectra}, respectively.

In \fig{fig: coherent 2d-spectra}, the spectrum is concentrated around $\kperp\rho = p\vth = 40$ at $t=0$ [see also Eq.\ \eqref{coherent init cond} and Table \ref{run table}].
The spectral density is transferred diagonally to the lower-left corner of the $(\kperp,p)$ space.
Since the high-$(\kperp,p)$ components suffer strong collisional dissipation, the upper-right energy content damps quickly.
The remaining lower-left transfer dominates after $t/\tau_{\rm init} \gtrsim 20$, and inverse cascade follows.

Nonlinear transfer of the invariants may be directly monitored as is done for the forward cascade simulation in Refs.\ \cite{Tatsuno-JPFRS10, NavarroCarati-arxiv10}.
Following Ref.\ \cite{AlexakisMininniPouquet-PRE05}, we define a shell filtered function by
\begin{align}
  \varphi_K(\Vec{r}) &:= \sum_{\Vec{k} \in {\cal K}} \phik e^{i\Vec{k}\cdot\Vec{r}},\\
  g_K(\Vec{R}) &:= \sum_{\Vec{k} \in {\cal K}} \gk e^{i\Vec{k}\cdot\Vec{R}},
\end{align}
where ${\cal K} = \{ \Vec{k}: K\rho-1/2 \leq |\Vec{k}|\rho < K\rho+1/2 \}$.
Then the evolution of energy in shell $K$ is described as
\begin{equation}
  \frac{d}{dt} \frac{n_0 q_i^2}{2T_{0i}}
    \sum_{\Vec{k} \in {\cal K}} (1-\varGamma_0) |\phik|^2
    = \sum_Q T^{(E)} (K,Q) - \mbox{collisions},
\end{equation}
where we introduced an energy transfer function
\begin{equation}
  T^{(E)}(K,Q) := - \frac{c q_i}{B_0 V} \iint \gyroavg{\varphi_K}{R}
    \{ \gyroavg{\varphi_Q}{R}, g \} \, d\Vec{R} \, d\Vec{v},
  \label{energy transfer function}
\end{equation}
which measures the rate of energy transferred from shell $Q$ to shell $K$.
Here $V$ denotes the spatial volume of the domain.
The entropy transfer function is defined in a similar manner in Ref.\ \cite{Tatsuno-JPFRS10} and is recaptured here in the present notation:
\begin{equation}
  T^{(W)}(K,Q) := - \frac{c q_i}{B_0 V} \iint \frac{g_K \{ \gyroavg{\varphi}{R},
    g_Q \} }{F_0} \, d\Vec{R} \, d\Vec{v}.
    \label{entropy transfer function}
\end{equation}
Note that the shell filtering is performed on $\varphi$ and $g$ in Eqs.\ \eqref{energy transfer function} and \eqref{entropy transfer function}, respectively, so that $T^{(E)}$ and $T^{(W)}$ both satisfy antisymmetry under exchange of $K$ and $Q$.

\begin{figure}
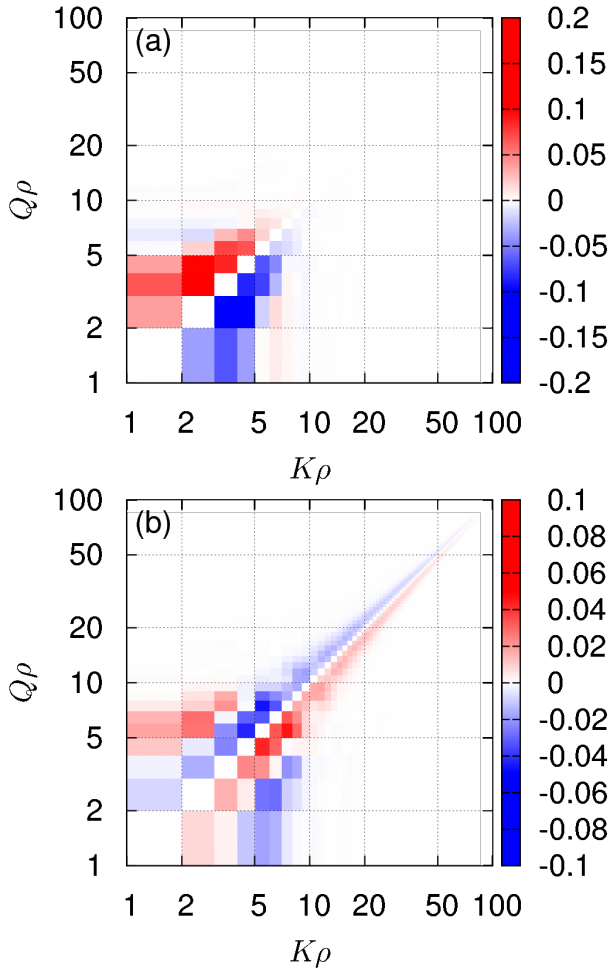
 
  \includegraphics[width=\figsize]{figs/esti20/etrans.esti20.t4.ps3}
  \includegraphics[width=\figsize]{figs/esti20/wtrans.esti20.t4.ps3}
  \caption{(Color online)
    Snapshot of the (a) energy transfer function $T^{(E)}(K,Q)/E$ and
    (b) entropy transfer function $T^{(W)}(K,Q)/W$ at $t/\tau_{\rm init} = 77$
    for Run D 
    [see Eqs.\ \eqref{energy transfer function},
    \eqref{entropy transfer function} and Table \ref{run table}].
  }
  \label{fig: coherent ktrans}
\end{figure}

Snapshots of the normalized energy transfer function $T^{(E)}(K,Q)/E$ and entropy transfer function $T^{(W)}(K,Q)/W$ at $t/\tau_{\rm init} = 77$ are shown for the simulation of coherent initial condition (Run D) in \fig{fig: coherent ktrans}. 
At this time the peak of the wave-number spectra (not shown) is located at $\kperp \rho \simeq 4$.
Figure \ref{fig: coherent ktrans}(a) shows that the energy transfer is (1) well localized along the diagonal (meaning local-scale interaction),
(2) has strong positive values at $(K\rho,Q\rho) = (3,4)$, $(2,4)$, $(2,3)$,
(3) has corresponding negative values at $(K\rho,Q\rho) = (4,3)$, $(4,2)$, $(3,2)$,
and (4) disappears rather quickly at high wave-number shells;
namely, the energy is transferred from large wave-number shells (small scales) to small wave-number shells (large scales) around the spectral peak, showing clear evidence of the inverse cascade of $E$.
This inverse transfer of $E$ creates the peaked, high-density region at the diagonal of $(\kperp,p)$ space, which propagates along the diagonal toward the small $(\kperp,p)$ regime as seen at $t/\tau_{\rm init}=19$ and $96$ of \fig{fig: coherent 2d-spectra}.
At an initial transient stage, however, we observe differences including nonlocal transfer, which is discussed elsewhere \cite{Plunk-Fjortoft, Plunk-Big-Fjortoft}.
Note that as the contribution to $E$ only comes from the $\kperp \rho = p \vth$ component of the distribution function, the transfer in velocity space proceeds in conjunction with that of position space, effectively ``unwinding'' fine structure in position and velocity space simultaneously.
This is a striking feature of the phase-space cascade.

On the other hand, from \fig{fig: coherent ktrans}(b), the entropy transfer (1) is well localized along the diagonal (meaning local-scale interaction),
(2) has a positive peak at $(K\rho,Q\rho) = (5,4)$ and corresponding negative peak at $(K\rho,Q\rho) = (4,5)$,
(3) extends to larger wave-number shells contrary to the transfer of $T^{(E)}$;
namely, the entropy is mostly transferred from small wave-number shells (large scales) to large wave-number shells (small scales), showing clear evidence of the forward cascade of $W$.
This forward transfer of $W$ creates the broad off-diagonal spectra seen at $t/\tau_{\rm init}=19$ and $96$ of \fig{fig: coherent 2d-spectra}, similar to the forward cascade simulation \cite{Tatsuno-PRL09,Tatsuno-JPFRS10} (see also \fig{fig: forward cascade result}).
However, it is interesting to note the reversed coloring around $(K\rho,Q\rho) \simeq (2,5)$ of \fig{fig: coherent ktrans}(b), which is located outside of the closest diagonal grids that show forward cascade of $W$.
This denotes the inverse transfer of $W$ associated with the strong inverse cascade of $E$; however, its magnitude is less than $1/3$ of the peak of the forward transfer of $W$ [note also the difference of the scale on the color bar in \fig{fig: coherent ktrans}(a) and (b)].

\begin{figure} 
  \centerline{
    \includegraphics[height=\fight,angle=270]{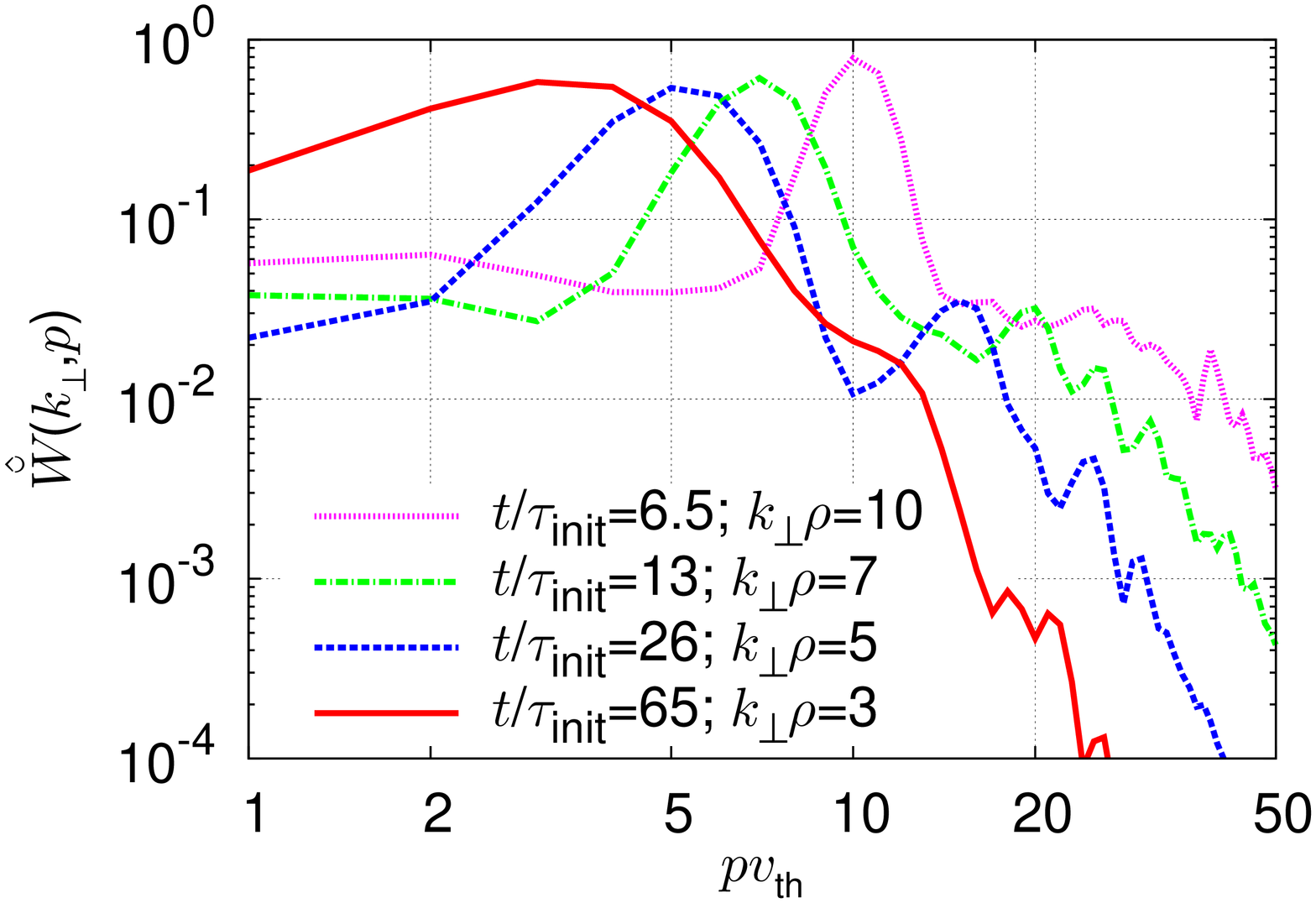}
  }
  \caption{(Color online)
    Slices of the 2D spectra of Run F along $p$ axis.
  }
  \label{fig: slices along p}
\end{figure}
In the case of random initial condition, the initial random velocity distribution \eqref{random init cond} is characterized by a vertical band in $(\kperp,p)$ space (see \fig{fig: random 2d-spectra}), covering $p \vth \lesssim 25$ (note that $k_0\rho =25$ as shown in Table \ref{run table}, Run F).
Most of the energy in the band is transferred to higher wave number and then dissipated by strong collisions as it is not associated with $E$ (recall the discussions in \sect{sec:phenomenology}).
For larger scales ($\kperp\rho < 25$), excitation of Fourier-Hankel modes is concentrated about the diagonal component ($\kperp \rho = p \vth$), as seen at $t/\tau_{\rm init} = 13$ and $65$.
As can be expected from the time evolution of the 2D spectra (compare Figs.\ \ref{fig: coherent 2d-spectra} and \ref{fig: random 2d-spectra}), random-initial-condition cases and the coherent-initial-condition cases show similar transfer after the transient phase (that is, the transfer functions resemble those of \fig{fig: coherent ktrans}).

Figure \ref{fig: slices along p} shows several slices of \fig{fig: random 2d-spectra}.
These slices are taken at the peak of the 2D spectra at each time.
Except at $t/\tau_{\rm init}=65$, the diagonal component $p\vth\simeq\kperp\rho$ is excited spontaneously with a nearly Gaussian form, which is in contrast to the initial $p$-spectrum consisting of a broad band occupying $p\vth \lesssim 25$.
Thus we conclude that the peaked excitation about the diagonal is not merely a reflection of a similarly peaked initial $p$-spectrum.
Furthermore, the peaking of the spectrum around $p\vth = \kperp\rho$, combined with Eq.\ \eqref{E W relationship}, justifies the approximation given in Eq.\ \eqref{invariants ratio}.
On the other hand, the Gaussian peak is surrounded by a fairly broad spectrum of an order of magnitude smaller amplitude, which is due to the excitation of random velocity fluctuation arising from the small-amplitude forward cascade [see \fig{fig: coherent ktrans}(b)].
Note that at $t/\tau_{\rm init}=65$, the spectrum has a fairly broad peak because the scale of the peak is approaching the system size (due to a background Maxwellian with the thermal velocity $\vth$).

In both cases, the spectra share some qualitative features with the runs reported in Ref.\ \cite{Tatsuno-JPFRS10} (see also \fig{fig: forward cascade result}):
At large $\kperp$ and $p$, the spectral density is broadly distributed about the diagonal, which is expected from the forward cascade of $W$ \cite{Tatsuno-JPFRS10}.
However, in each case there is a highly peaked diagonal component (whose scale is denoted by $l_*$ in \sect{sec:decay law}) that tends to be the dominant contribution to the total value of the invariants at the later stage.
This is the component generated by the inverse cascade and is discussed in more detail in the following sections.

\subsection{Decay laws}
\label{sec: num decay law}

\begin{figure} 
  \centerline{
    \includegraphics[height=\fight,angle=270]
                    {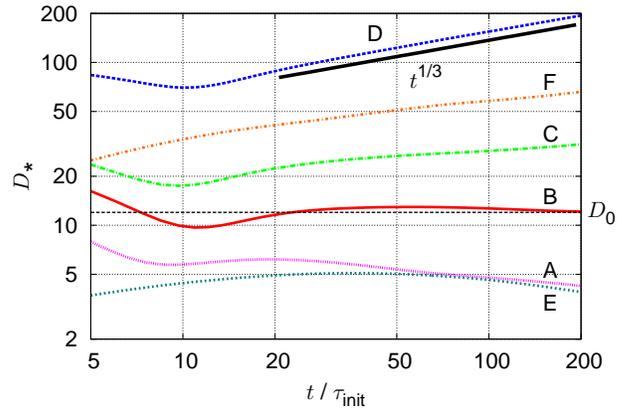}
  }
  \caption{(Color online)
    Time evolution of $D_*$ for the runs indexed
    in Table \ref{run table}.
    Runs A and E correspond to the strongly collisional case, B corresponds to
    the marginal case and C, D and F correspond to the weakly collisional case.
  }
  \label{fig: dorland number}
\end{figure}

The time evolution of the dimensionless number $D_*$ is shown for each of the runs in \fig{fig: dorland number}.
Depending on the long-time behavior, we may classify the runs into strongly collisional, marginal and weakly collisional cases as described in \sect{sec:decay law}.
Run B corresponds to the marginal case as $D_*$ approaches a constant ($D_0 \simeq 12$) in $t/\tau_{\rm init} \gtrsim 30$.
Runs A and E are strongly collisional as $D_*$ decreases in time and Runs C, D and F are weakly collisional as $D_*$ increases in time.
The evolution of $D_*$ differs among weakly collisional runs but approaches the theoretical limit $D_* \propto t^{1/3}$ as $D_*$ increases.

\begin{figure} 
  \centerline{
    \includegraphics[height=\fight,angle=270]{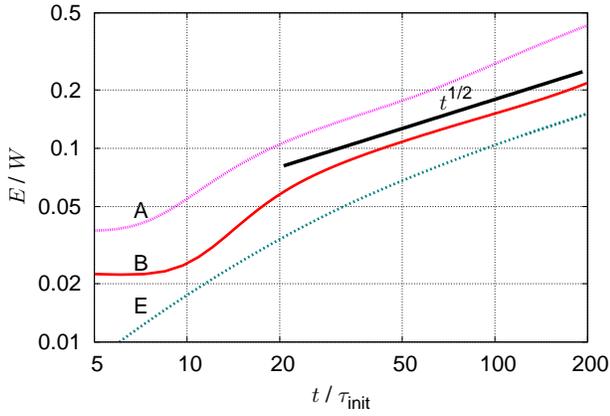}
  }
  \caption{(Color online)
    Time evolution of the ratio of collisionless invariants for the
    strongly collisional (Runs A and E) and marginal (Run B) cases.
    The decay law \eqref{collisional decay law} is drawn for comparison.
  }
  \label{fig: ratio law}
\end{figure}

Figure \ref{fig: ratio law} shows the time evolution of the ratio of two collisionless invariants for strongly collisional (Runs A and E) and marginal (Run B) cases.
Initially, the ratio is of the order of $1/(k_0\rho)$ or $1/(k_0^2 \rho^2)$ depending on the initial velocity distributions [see Eqs.\ \eqref{coherent init ratio} and \eqref{random init ratio}].
Coherent cases show an initial phase of constant ratio ($t/\tau_{\rm init} \lesssim 6$ for Run A and $t/\tau_{\rm init} \lesssim 8$ for Run B), which stems from the fact that the initial condition \eqref{coherent init cond} is almost monochromatic with high-$(\kperp,p)$, and that both collisionless invariants initially decay at the same rate due to the collisional damping of the distribution function at the scale $k_0^{-1}$.
The decay law \eqref{collisional decay law} seems consistent with both strongly collisional and marginal cases at the later stage.
The case with the random initial condition tends to take a longer time to approach the theoretical line as the initial ratio, $E / W \sim 1 / (k_0^2 \rho^2)$ [see Eq.\ \eqref{random init ratio}], is much smaller than the asymptotic ratio $1 / (k_*\rho)$.
A slight deviation from the power law is observed at the last stage of the simulation for Runs A and B, which is due to the fact that the cascade has reached the largest wave length of the system around $E / W \gtrsim 0.2$, due to the finite size of the simulation box.

\begin{figure} 
  \centerline{
    \includegraphics[height=\fight,angle=270]{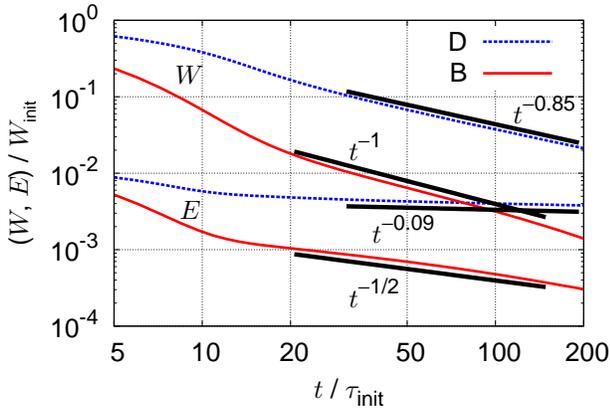}
  }
  \caption{(Color online)
    Time evolution of the invariants for the marginal (Run B) and weakly
    collisional (Run D) cases.
    Decay law of each invariant \eqref{marginal decay law} are drawn for the
    marginal case, and fitted slopes are drawn for the weakly collisional case.
  }
  \label{fig: decay law}
\end{figure}

The time evolution of individual collisionless invariants are shown in \fig{fig: decay law} for the marginal (Run B) and weakly collisional (Run D) cases.
The marginal run (B) shows a reasonable agreement with the theoretical expectation \eqref{marginal decay law}.
The weakly collisional run shows a significantly slower decay than the marginal case, but still faster than the asymptotic ($\nu \to 0$) limit \eqref{asymptotic decay law}.
Computational resources beyond those available for this study will be needed to unequivocally confirm the asymptotic decay laws \eqref{asymptotic decay law}.

\subsection{Self-similarity of spectra}

The self-similarity of the spectrum may be investigated with the approach of Chasnov for 2D fluid turbulence \cite{Chasnov-PoF97}.
Chasnov defined an instantaneous length scale of the decaying turbulence in terms of the ratio of the two invariants, energy $\langle \Vec{u}^2 \rangle$ and enstrophy $\langle \omega^2 \rangle$.
This is possible because of the relationship that hold between them at each scale, namely vorticity $\omega$ is a first-order derivative of velocity $\Vec{u}$.
Here we may define an instantaneous length scale in terms of the ratio of the two collisionless invariants $W$ and $E$, based on Eq.\ \eqref{E W relationship}.
In general $W$ and $E$ can be nearly independent due to the extra degree of freedom arising from the velocity space; however, the present argument is valid when the spectral density is sufficiently concentrated along the diagonal, $\kperp\rho = p\vth$.

We normalize the spectra in velocity space as well as in position space.
On dimensional grounds, we may define
\begin{align}
  \Norm{W}_k(\Norm{k}) &= \frac{k_* \Four{W}(\kperp,t)}{W},
  \label{eq: normalized spectra: Egk}\\
  \Norm{E}_k(\Norm{k}) &= \frac{k_* \Four{E}(\kperp,t)}{E},\\
  \label{eq: normalized spectra: Ephi}
  \Norm{W}_p(\Norm{p}) &= \frac{p_* \Hank{W}(p,t)}{W},
\end{align}
where $k_*$ and $p_*$ are the inverses of the characteristic scale length in position and velocity space, respectively, determined by [see Eq.\ \eqref{invariants ratio}],
\begin{align}
  &k_* = \frac{W}{\rho E},
  &p_* = \frac{W}{\vth E},
  \label{characteristic length}
\end{align}
the wave number $\kperp$ and velocity wave number $p$ are normalized to $k_*$ and $p_*$,
\begin{align}
  &\Norm{k} = \frac{\kperp}{k_*},
  &\Norm{p} = \frac{p}{p_*},
\end{align}
and the wave-number and velocity-space spectra are defined by
\begin{align}
  \Four{W}(\kperp,t) &= \sum_{|\Vec{k}| = \kperp}
    \int \frac{T_{0i} |\gk|^2}{2 F_0} \, d\Vec{v}, \\
  \Hank{W}(p,t) &= \sum_{\Vec{k}} p |\FourHank{g}(\Vec{k},p)|^2,
\end{align}
and Eq.\ \eqref{def: E spectrum}.

When we apply the normalization \eqref{eq: normalized spectra: Egk}--\eqref{eq: normalized spectra: Ephi} to simulation results, one would expect a good coincidence for the same value of $D_*$.
Namely, if the theory is correct, the normalized spectra at different times should collapse in the time-asymptotic regime of the marginal case (Run B, see also \fig{fig: dorland number}).

\begin{figure} 
  \includegraphics[height=\figsize,angle=270]{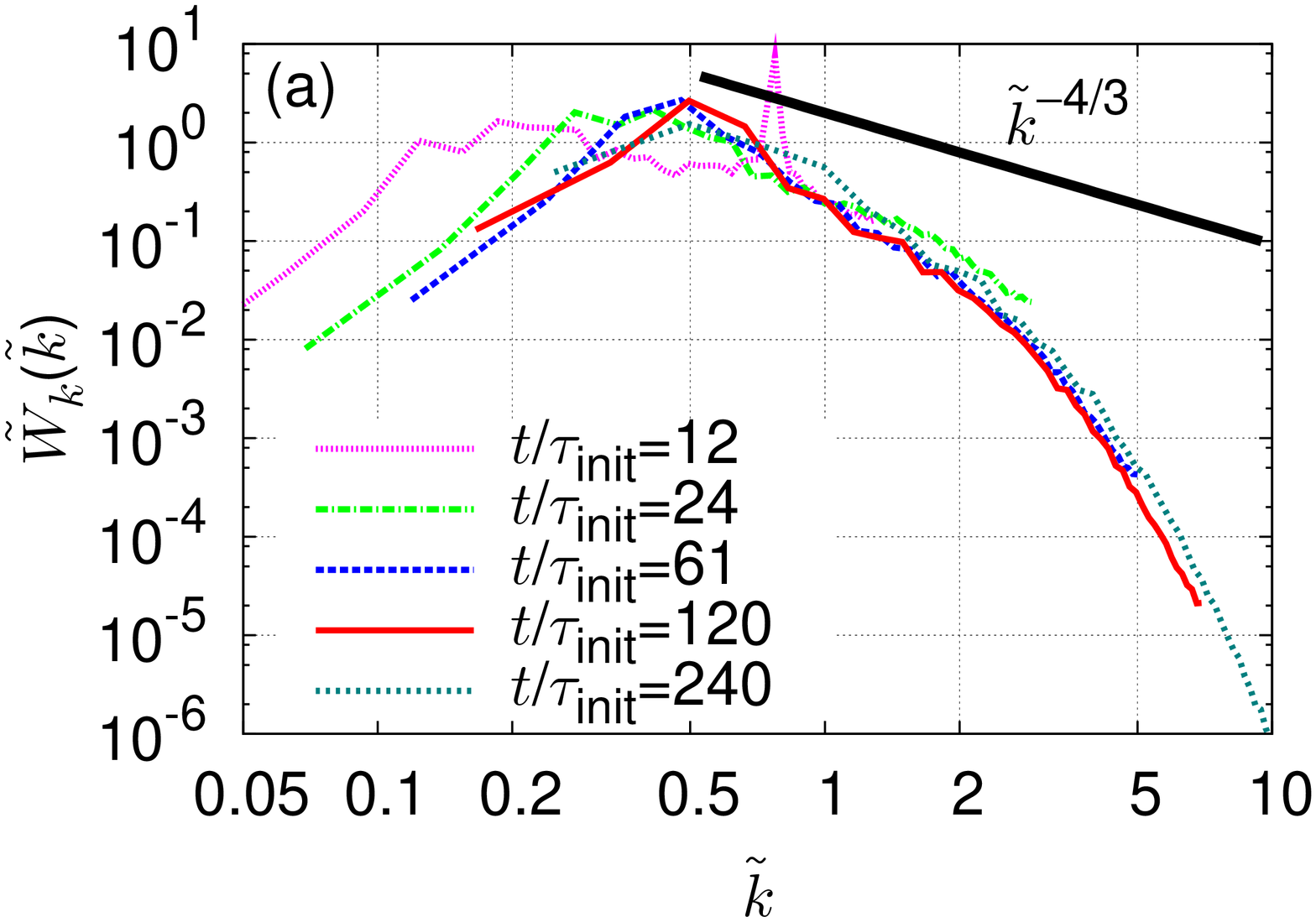}
  \includegraphics[height=\figsize,angle=270]{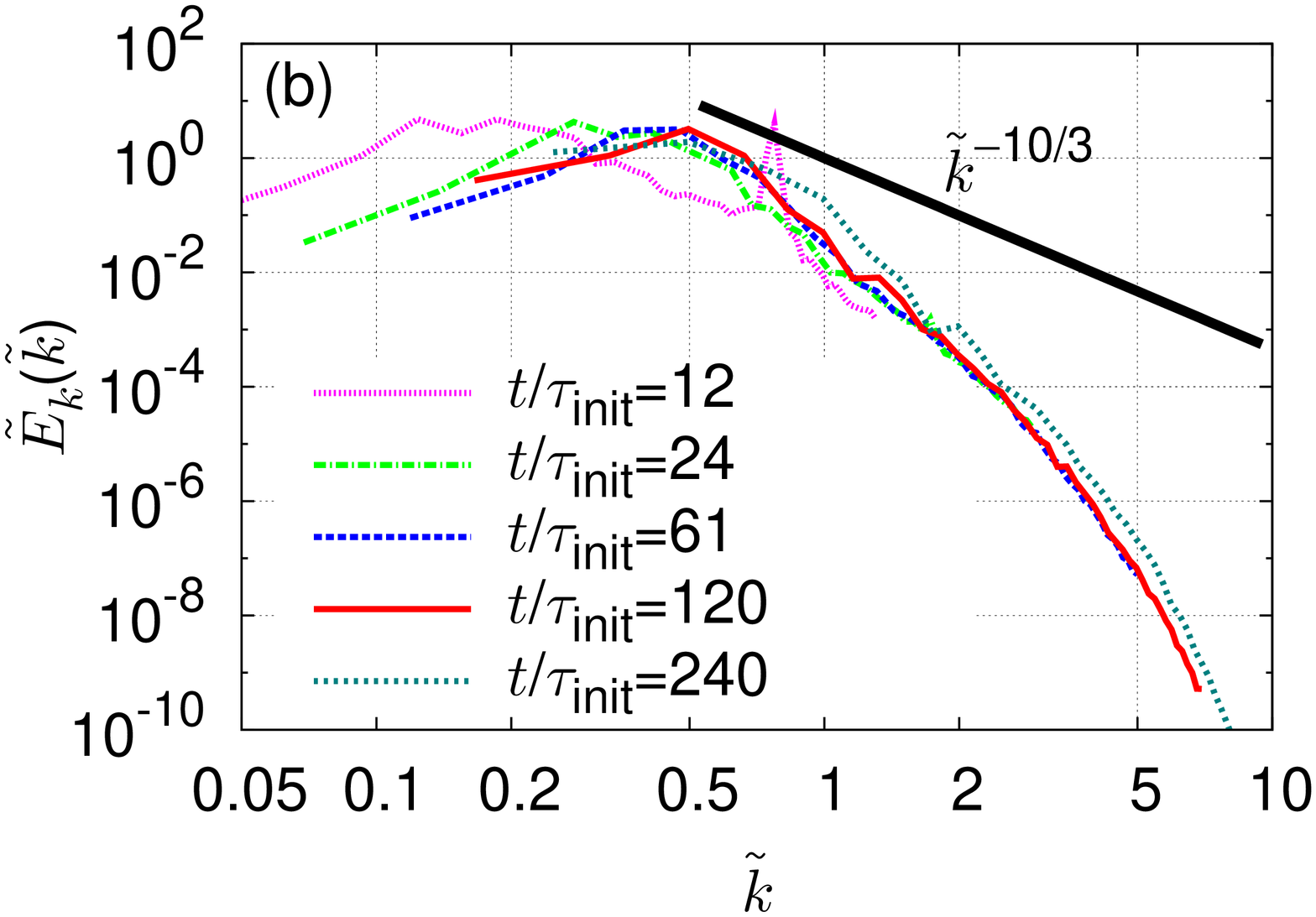}
  \includegraphics[height=\figsize,angle=270]{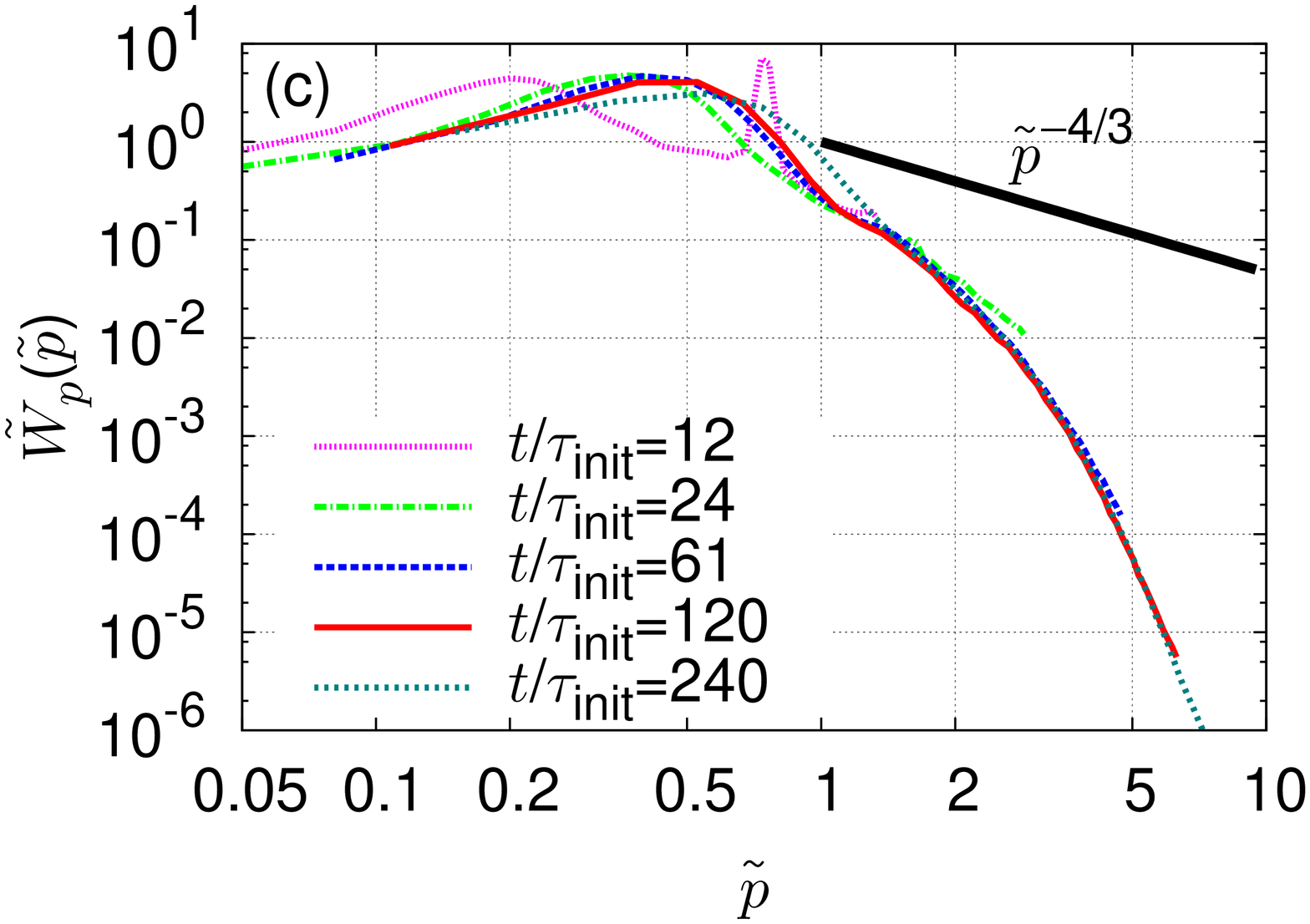}
  \caption{(Color online)
    Normalized spectra for Run B (see Table \ref{run table}).
  }
  \label{fig: collapsed spectra for Run B}
\end{figure}

We first show the normalized spectra defined by Eqs.\ \eqref{eq: normalized spectra: Egk}--\eqref{eq: normalized spectra: Ephi} in \fig{fig: collapsed spectra for Run B} for Run B (see Table \ref{run table}).
As one can easily expect from the 2D spectra (\fig{fig: coherent 2d-spectra}) and the ratio (\fig{fig: ratio law}), the spectral peak moves towards larger scales [to smaller $(\kperp,p)$] and we may expect self-similar spectra in the later stage of the simulation (From \figs{fig: dorland number}{fig: decay law}, we see that the asymptotic regime is attained in the range $30 \lesssim t/\tau_{\rm init} \lesssim 150$).
After the ratio $E/W$ approaches the power-law behavior ($t/\tau_{\rm init} \gtrsim 50$, see \fig{fig: ratio law}), wave-number spectra show promising coincidence which indicates a good self-similarity of the decaying turbulence.
Notice especially the amazing coincidence of the tail at $t/\tau_{\rm init}=61$ and $120$, both in the time-asymptotic regime as shown in \figs{fig: dorland number}{fig: decay law}.
We also confirm that the peak of the spectra moves towards large scales roughly with the scaling $k_* \propto t^{-1/2}$ in this time regime.
The last one ($t/\tau_{\rm init} = 240$) is offset by a small amount because of the fact that the spectral peak has reached the system size.
The wave-number spectra at the right of the peak show slopes somewhat steeper than the theoretical prediction of the forward cascade \cite{Tatsuno-PRL09, Tatsuno-JPFRS10}.
This agrees with the expectation from 2D fluid turbulence \cite{Chasnov-PoF97}, as the dimensionless number $D_*$ is not asymptotically large in the marginal case.

\begin{figure} 
  \includegraphics[height=\figsize,angle=270]{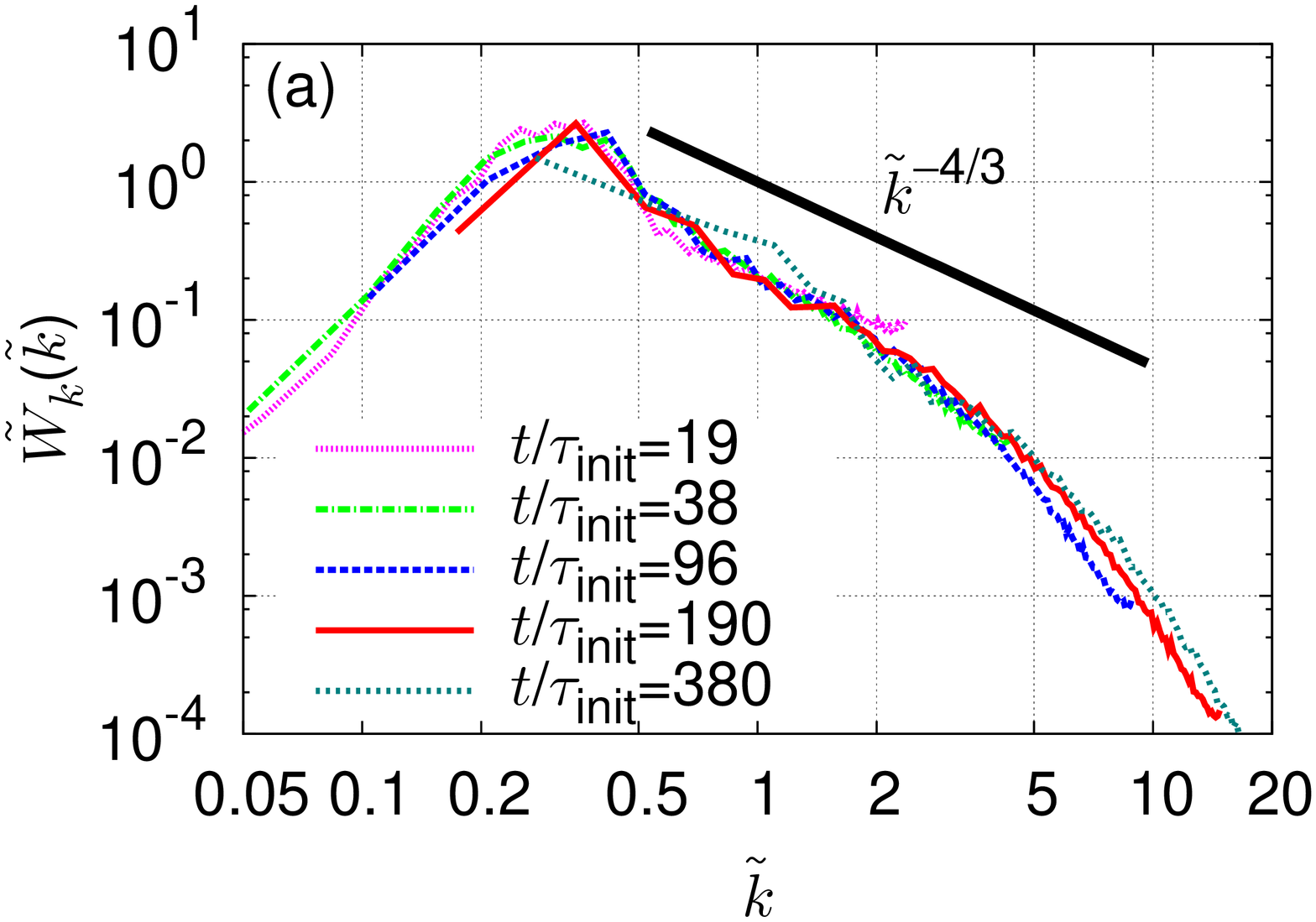}
  \includegraphics[height=\figsize,angle=270]{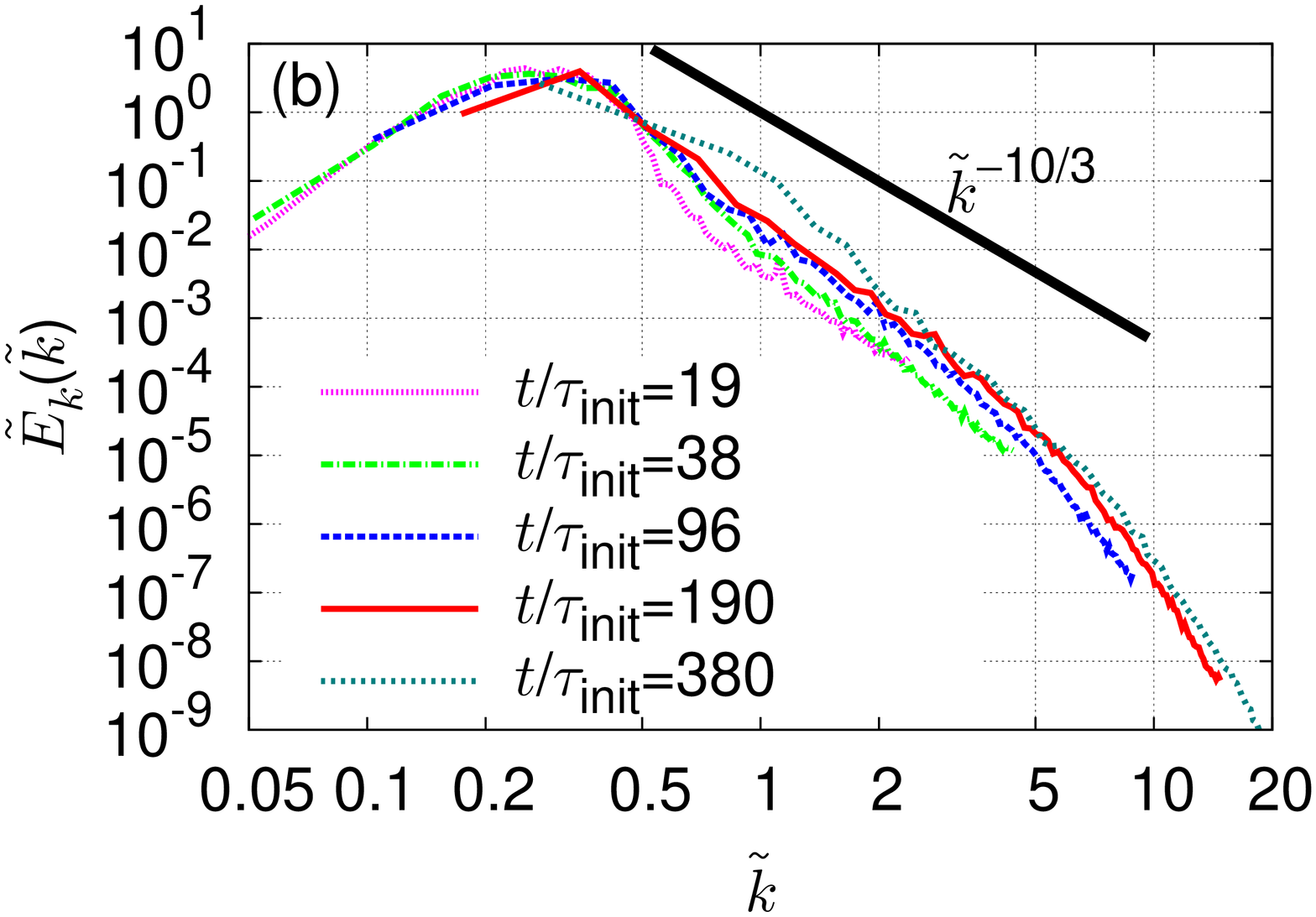}
  \includegraphics[height=\figsize,angle=270]{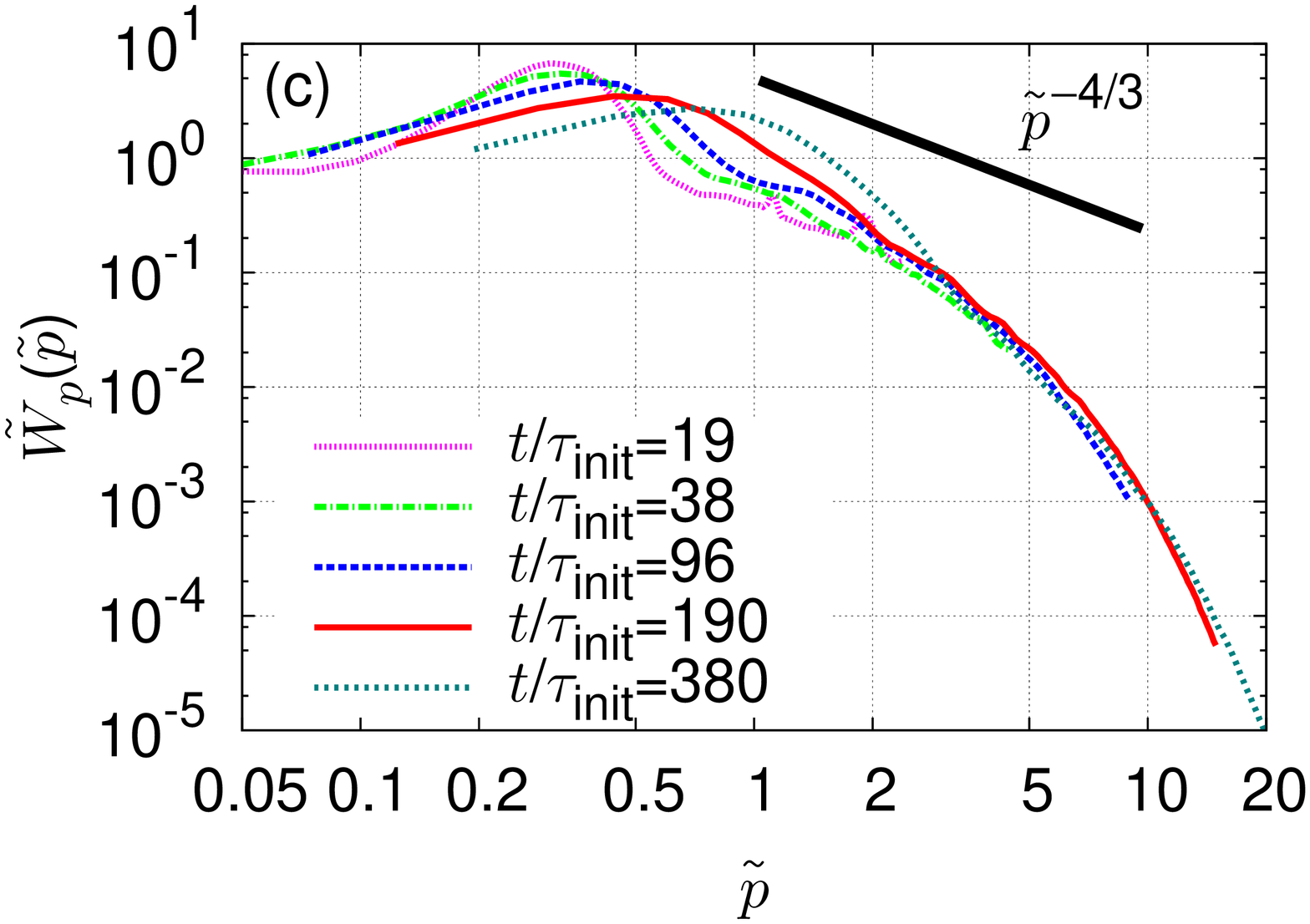}
  \caption{(Color online)
    Normalized spectra for Run D (see Table \ref{run table}).
  }
  \label{fig: collapsed spectra for Run D}
\end{figure}

As $D_*$ becomes large, the spectral slope becomes shallower and the forward cascade spectra with the slope $-4/3$ for $\Four{W}$ and $-10/3$ for $\Four{E}$ (see Ref.\ \cite{Tatsuno-PRL09}) is expected to be recovered.
Figure \ref{fig: collapsed spectra for Run D} shows the normalized spectra for Run D.
Comparison with \fig{fig: collapsed spectra for Run B} clearly indicates that the spectral slopes in \fig{fig: collapsed spectra for Run D} are indeed closer to the theoretical expectation of the forward cascade, for both spectra of $\Norm{W}_k$ and $\Norm{E}_k$.
Collapse of the spectra is as good as \fig{fig: collapsed spectra for Run B}.
However, as $D_*$ continues to grow for $t/\tau_{\rm init} \gtrsim 30$ (see \fig{fig: dorland number}), the spectral slope does change slightly, gradually approaching the theoretical prediction ($\Norm{W}_k \sim \Norm{k}^{-4/3}$ and $\Norm{E}_k \sim \Norm{k}^{-10/3}$) as shown in \fig{fig: collapsed spectra for Run D}.
At $t/\tau_{\rm init} = 380$ the spectral peak has reached the system size and the spectra become offset.

\section{Summary}
\label{sec:summary}

We presented theoretical and numerical investigations of electrostatic, freely decaying turbulence of weakly-collisional, magnetized plasmas using the gyrokinetic model in 4D phase space (two position-space and two velocity-space dimensions).
Landau damping was removed from the system by ignoring variation along the background magnetic field.
Nonlinear interactions introduce an amplitude-dependent perpendicular phase mixing of the gyrophase-independent part of the perturbed distribution function, which creates structure in $v_\perp$ comparable in size to spatial structure (see \fig{fig:mixing}).

Since our 2D (in position space) system possesses two collisionless invariants [entropy, or free energy, see Eq.\ \eqref{eq:wg}; and energy, see Eq.\ \eqref{eq:esi}], a dual cascade (forward and inverse cascades) takes place when the initial condition consisted of small-scale fluctuations in position as well as in velocity space such as in Eqs.\ \eqref{coherent init cond} and \eqref{random init cond}.
As the dual cascade proceeds, the peak of the spectra moves towards large scales in both position and velocity space as shown in \figs{fig: coherent 2d-spectra}{fig: random 2d-spectra}.
Nonlinear transfer is diagnosed by the direct numerical simulation, which shows a clear evidence of inverse (forward) transfer of energy (entropy) (see \fig{fig: coherent ktrans}).
In the inverse cascade, the velocity space spectrum is highly focused due to the fact that energy comes from coherent structure in the velocity space [see Eq.\ \eqref{E W relationship} and \fig{fig: slices along p}], which is in contrast to the broad distribution of velocity scales excited at each wave number in the forward cascade \cite{Tatsuno-PRL09, Tatsuno-JPFRS10}.

Following an example from 2D Navier-Stokes turbulence \cite{Chasnov-PoF97}, a phenomenological theory of decay is presented (\sect{sec:decay law}) as well as the numerical simulation (\sect{sec: num decay law}).
Several types of asymptotic decay have been identified in numerical simulation, which match up well with the phenomenological theory using a classification based on the kinetic dimensionless number $D_*$ [see Eq.\ \eqref{microscopic Dorland number}].
When $D_*$ takes a marginal value, decay laws of both invariants are identified [see Eq.\ \eqref{marginal decay law}, Figs.\ \ref{fig: dorland number} and \ref{fig: decay law}].
In the weakly collisional regime the invariants decay more slowly [see \fig{fig: decay law}]; and in the asymptotic limit where the collision frequency becomes negligible (but finite), the entropy (or free energy) decays as $t^{-2/3}$ while the energy stays constant [see Eq.\ \eqref{asymptotic decay law}].

In this paper we focused on the time-asymptotic regime of the freely decaying turbulence.
Although there is a range of behavior depending on the strength of dissipation, the cases are unified by some common features.
The most prominent of these features is the dual cascade, whereby the invariant $E$ cascades inversely to large scales while the invariant $W$ cascades to small scales.
The transient and driven cases, hosts to a broader range of phenomena, have recently been explored in other works \cite{Plunk-Fjortoft, Plunk-Big-Fjortoft}.

\begin{acknowledgments}
  Numerical simulations were performed at NERSC, NCSA, TACC, and IFERC CSC.
  The authors thank Drs.\ A. Schekochihin and S. Cowley for fruitful discussions.
  This work was supported by the U.S. DOE Center for Multiscale Plasma Dynamics, the Leverhulme Trust International Network for Magnetised Plasma Turbulence, Wolfgang Pauli Institute, and JSPS KAKENHI Grant Number 24540533.
\end{acknowledgments}

\appendix

\section{Collision operator}
\label{sec: collision}

The collision operator that we use in our numerical simulations is a linearized model operator described in Refs.\ \cite{Abel-PoP08,Barnes-PoP09}.
It consists of pitch-angle scattering and energy diffusion, augmented by a few correction terms introduced for conservation of momentum and energy.
Symbolically it is written as
\begin{equation}
  C = L + D + U_L + U_D + E,
\end{equation}
where $L$ and $D$ denotes the diffusion-type operator in pitch angle and energy, respectively, $U_L$ and $U_D$ are momentum restoring terms associated with $L$ and $D$, respectively, and $E$ is the energy restoring term associated with $D$.

Specifically,
\begin{equation}
  L = \frac{\nu_D}{2} \frac{\partial}{\partial \xi} \left[ (1-\xi^2)
  \frac{\partial}{\partial \xi} \right],
\end{equation}
and
\begin{equation}
  D = \frac{1}{2v^2} \frac{\partial}{\partial v} \left( \nu_{\parallel} v^4 F_0
    \frac{\partial}{\partial v} \frac{1}{F_0} \right),
\end{equation}
where $\nu_D$ and $\nu_{\parallel}$ are collision frequencies and $\xi=\vpar/v$ is the pitch angle.
It is noted that $L$ handles the smoothing in $\xi$ space while $D$ does it for energy, so with both of them we can cover the smoothing in the whole two-dimensional plane of $\vperp$ and $\vpar$.
The detailed expression of collision frequencies and conservation terms are shown in Refs.\ \cite{Abel-PoP08,Barnes-PoP09}.


\begin{thebibliography}{99}
\bibitem{4D simulation}
  Essential features of the collisionless dynamics may be described in a 3D phase space $(x, y, \vperp)$, however, we use four phase-space dimensions $(x, y, \vperp, \vpar)$ to account for the fact that collisions isotropize the velocity space.
\bibitem{Hankel transform}
  Here the Hankel transform is defined by the three-dimensional velocity
  integral.  When there is no dependence on gyroangle and Maxwellian dependence
  on $\vpar$, we will obtain the standard Hankel transform with respect to
  $\vperp$ times $2 \pi$.
\bibitem{dstar}
  $D_*$ is the local version of $D$ of Ref.\ \cite{Tatsuno-PRL09}: As the latter was defined on the Larmor scale, i.e., $D$ is the same as $D_*$ for $k_*\rho=1$.
\end{thebibliography}
\end{document}